\documentclass[lettersize,journal,hidelinks]{IEEEtran}
\usepackage{amsmath,amssymb,amsfonts}
\usepackage{algorithmic}
\usepackage{algorithm}
\usepackage{array}
\usepackage{textcomp}
\usepackage{stfloats}
\usepackage{url}
\usepackage{verbatim}
\usepackage{graphicx}
\usepackage{cite}
\usepackage{lipsum}
\usepackage[acronym,shortcuts]{glossaries}
\usepackage[dvipsnames]{xcolor}
\usepackage{bm}
\usepackage{caption}
\usepackage{subcaption}
\usepackage{relsize}
\usepackage{soul}
\usepackage{outlines}
\usepackage{isomath}
\usepackage{orcidlink}
\usepackage[bottom]{footmisc}
\usepackage{comment}
\usepackage{mleftright}
\usepackage{orcidlink}
\usepackage{multirow} 
\renewenvironment{IEEEbiography}[1]
{\IEEEbiographynophoto{#1}}
{\endIEEEbiographynophoto}

\newacronym{V2X}{V2X}{vehicle-to-everything}
\newacronym{V2V}{V2V}{vehicle-to-vehicle}

\newacronym{OS-QSM}{OS-QSM}{optimized scalable QSM}
\newacronym{GSM}{GSM}{generalized spatial modulation}
\newacronym{FDFR}{FDFR}{full-diversity full-rate}
\newacronym{mMIMO}{mMIMO}{massive multiple-input multiple-output}
\newacronym{MIMO}{MIMO}{multiple-input multiple-output}
\newacronym{MU}{MU}{multi-user}
\newacronym{OFDM}{OFDM}{orthogonal frequency-domain multiplexing}
\newacronym{IM}{IM}{index modulation}
\newacronym{IoT}{IoT}{Internet-of-Things}
\newacronym{QSM}{QSM}{quadrature spatial modulation}
\newacronym{BP}{BP}{belief propagation}
\newacronym{GaBP}{GaBP}{Gaussian belief propagation}
\newacronym{SM}{SM}{spatial modulation}
\newacronym{IQ}{IQ}{in-phase and quadrature}
\newacronym{ML}{ML}{machine learning}
\newacronym{BER}{BER}{bit error rate}
\newacronym{P2P}{P2P}{point-to-point}
\newacronym{AWGN}{AWGN}{additive white Gaussian noise}
\newacronym{SWIPT}{SWIPT}{simultaneous wireless information and power transfer}
\newacronym{CSI}{CSI}{channel state information}
\newacronym{MQAM}{$M$-QAM}{$M$-ary quadrature amplitude modulation}
\newacronym{IC}{IC}{interference cancellation}
\newacronym{SGA}{SGA}{scalar Gaussian approximation}
\newacronym{CLT}{CLT}{central limit theorem}
\newacronym{PDF}{PDF}{probability density function}
\newacronym{GB-ISTA}{GB-ISTA}{greedy boxed iterative soft-thresholding algorithm}
\newacronym{MP}{MP}{message passing}
\newacronym{WL}{WL}{wireless localization}
\newacronym{SD}{SD}{sphere decoder}
\newacronym{FD}{FD}{full-duplex}
\newacronym{STC}{STC}{space-time coding}
\newacronym{SotA}{SotA}{state-of-the-art}
\newacronym{IER}{IER}{index vector error rate}
\newacronym{B5G}{B5G}{beyond fifth generation}
\newacronym{STC-SM}{STC-SM}{space-time coded SM}
\newacronym{STC-QSM}{STC-QSM}{space-time coded QSM}
\newacronym{SC-IM}{SC-IM}{single-carrier IM}
\newacronym{SSK}{SSK}{space shift keying}
\newacronym{mmWave}{mmWave}{millimeter-wave}
\newacronym{THz}{THz}{Terahertz}
\newacronym{RIS}{RIS}{reflective intelligence surface}
\newacronym{RF}{RF}{radio frequency}
\newacronym{STBC}{STBC}{space-time block code}
\newacronym{MMSE}{MMSE}{minimum mean-squared-error}
\newacronym{CS}{CS}{compressive sensing}
\newacronym{i.i.d.}{i.i.d.}{independent and identically distributed}
\newacronym{JCAS}{JCAS}{joint communication and sensing}
\newacronym{ISAC}{ISAC}{integrated sensing and communication}
\newacronym{JRC}{JRC}{joint radar-communications}
\newacronym{SE}{SE}{spectral efficiency}
\newacronym{5G}{5G}{fifth generation}
\newacronym{EE}{EE}{energy efficiency}
\newacronym{RBL}{RBL}{rigid body localization}
\newacronym{RBT}{RBT}{rigid body tracking}
\newacronym{SC-RBL}{SC-RBL}{soft-connected RBL}
\newacronym{W-RBL}{W-RBL}{\underline{wireless} RBL}
\newacronym{GA}{GA}{genie-aided}
\newacronym{MC}{MC}{matrix completion}
\newacronym{EA}{EA}{``\emph{estimate-then-average}''}
\newacronym{AE}{AE}{``\emph{average-then-estimate}''}
\newacronym{IRS}{IRS}{intelligent reflecting surface}
\newacronym{RSSI}{RSSI}{received signal strength indicator}
\newacronym{PSO}{PSO}{particle swarm optimization}

\newacronym{NTN}{NTN}{non-terrestrial networks} 
\newacronym{6G}{6G}{sixth-generation}
\newacronym{3D}{3D}{three-dimensional}
\newacronym{D2D}{D2D}{device-to-device}
\newacronym{RR}{RR}{round-robin}
\newacronym{DA}{DA}{Dutch auction}
\newacronym{CWFL}{CWFL}{clustered WFL}
\newacronym{WFL}{WFL}{wireless federated learning}
\newacronym{RSMA}{RSMA}{rate splitting multiple access}

\newacronym{TDMA}{TDMA}{time-domain multiple access}
\newacronym{NOMA}{NOMA}{non-orthogonal multiple access}

\newacronym{CT}{CT}{compute-then-transmit}
\newacronym{SDP}{SDP}{semidefinite programming}
\newacronym{FP}{FP}{fractional programming}
\newacronym{CF-mMIMO}{CF-mMIMO}{cell free massive MIMO}
\newacronym{iid}{i.i.d.}{independent and identically distributed}
\newacronym{DL}{DL}{downlink}
\newacronym{UL}{UL}{uplink}
\newacronym{MDS}{MDS}{multidimensional scaling}

\newacronym{SIC}{SIC}{successive interference cancellation}
\newacronym{BS}{BS}{base station}
\newacronym{TX}{TX}{transmit}
\newacronym{RX}{RX}{receive}

\newacronym{SISO}{SISO}{single-input single-output}

\newacronym{SINR}{SINR}{signal-to-interference-and-noise ratio}
\newacronym{FL}{FL}{federated learning}
\newacronym{CPU}{CPU}{central processing unit}
\newacronym{KNN}{KNN}{K-nearest-neighbor}

\newacronym{GD}{GD}{gradient descent}

\newacronym{RSS}{RSS}{received signal strength}
\newacronym{FIM}{FIM}{fisher information matrix}
\newacronym{ToA}{ToA}{time of arrival}
\newacronym{AoA}{AoA}{angle of arrival}
\newacronym{ADoA}{ADoA}{angle difference of arrival}
\newacronym{GP}{GP}{Gaussian process}
\newacronym{2D}{2D}{two-dimensional}
\newacronym{GPR}{GPR}{Gaussian process regression}
\newacronym{GNSS}{GNSS}{global navigation satellite systems}

\newacronym{RRH}{RRH}{remote radio head}
\newacronym{GPS}{GPS}{Global Positioning System}
\newacronym{RFID}{RFID}{radio frequency identification}
\newacronym{TCAS}{TCAS}{traffic alert and collision avoidance systems}
\newacronym{RMSE}{RMSE}{root mean square error}
\newacronym{MSE}{MSE}{mean square error}
\newacronym{SGD}{SGD}{stochastic gradient descent}

\newacronym{CU}{CU}{computing unit}
\newacronym{DM-MIMO}{DM-MIMO}{distributed massive multiple-input multiple-output}
\newacronym{LOS}{LOS}{line-of-sight}
\newacronym{NLOS}{NLOS}{non-line-of-sight}
\newacronym{ROI}{ROI}{region of interest}
\newacronym{AP}{AP}{access point}
\newacronym{PoCs}{PoCs}{projections onto convex sets}
\newacronym{TDOA}{TDOA}{time difference of arrival}
\newacronym{DoA}{DoA}{direction of arrival}
\newacronym{UE}{UE}{user equipment}
\newacronym{dB}{dB}{decibel}

\newacronym{CG}{CG}{conjugate gradient}
\newacronym{SC}{SC}{soft-connected}
\newacronym{CRLB}{CRLB}{Cramér-Rao Lower Bound}
\newacronym{PoA}{PoA}{phase of arrival}
\newacronym{UAV}{UAV}{unmanned aerial vehicle}
\newacronym{VR}{VR}{virtual reality}
\newacronym{ITS}{ITS}{intelligent transportation system}

\newacronym{SLAM}{SLAM}{simultaneous localization and mapping}
\newacronym{WLS}{WLS}{weighted least square}
\newacronym{JSCC}{JSCC}{joint sensing communication and computing}
\newacronym{SMDS}{SMDS}{super multidimensional scaling}
\newacronym{EDM}{EDM}{euclidean distance matrix}
\newacronym{LDPC}{LDPC}{low density parity check}
\newacronym{MAP}{MAP}{maximum a posteriori}

\newcommand{\tablecheck}{\raisebox{-0.55em}{\Large\checkmark}}

\begin{document}


\title{ Rigid Body Localization and Tracking for 6G V2X: \\Algorithms, Applications, and Road to Adoption\vspace{-.5ex}}

\author{Niclas~F\"uhrling\textsuperscript{\orcidlink{0000-0003-1942-8691}}, \IEEEmembership{Graduate Student Member, IEEE}, 
\,Hyeon~Seok~Rou\textsuperscript{\orcidlink{0000-0003-3483-7629}}, \IEEEmembership{Member, IEEE},\\
Giuseppe~Thadeu~Freitas~de~Abreu\textsuperscript{\orcidlink{0000-0002-5018-8174}}, \IEEEmembership{Senior Member, IEEE}, 
\,David~Gonz{\'a}lez~G.\textsuperscript{\orcidlink{0000-0003-2090-8481}},  \IEEEmembership{Senior Member, IEEE},\\
Gonzalo Seco-Granados\textsuperscript{\orcidlink{0000-0003-2494-6872}}\IEEEmembership{Senior Member, IEEE},
Osvaldo~Gonsa\textsuperscript{\orcidlink{0000-0001-5452-8159}}.\vspace{-2ex}

\vspace{-4ex}

\thanks{N.~F\"uhrling, H.~S.~Rou and G.~T.~F.~Abreu are with the School of Computer Science and Engineering, Constructor University, Campus Ring 1, 28759, Bremen, Germany (emails: [nfuehrling, hrou, gabreu]@constructor.university).}
\thanks{D.~Gonz{\'a}lez~G. and O.~Gonsa are with the Wireless Communications Technologies Group, Continental Automotive Technologies GmbH, Guerickestrasse 7, 60488, Frankfurt am Main, Germany (emails: david.gonzalez.g@ieee.org, osvaldo.gonsa@continental-corporation.com).}
\thanks{G.~Seco-Granados is with the Department of Telecommunications and Systems Engineering, Universitat Autònoma de Barcelona, Spain (email: gonzalo.seco@uab.cat).}
}

\markboth{Submitted to the IEEE Vehicular Technology Magazine Special Issue}%
{N.~F\"{u}hrling \MakeLowercase{\textit{et al.}}}

\maketitle

\begin{abstract}
\Ac{V2X} perception refers to a suite of technologies that empower vehicles to sense their environment and communicate with other entities, including surrounding vehicles, infrastructure, and cloud/edge networks. 
With the growing demands of autonomous driving, \ac{V2X} perception has gained significant attention, particularly through the emergence of \ac{ISAC} frameworks. 
Within this landscape, \ac{RBL} has emerged as a promising paradigm, enabling the estimation of not only the position and velocity of the targets, but also its \ac{3D} geometric structure and orientation. 
This article introduces the concept of \ac{RBL}, highlights its unique advantages and applications, identifies key technical challenges, and finally outlines future research directions. 
In addition, the potential of \ac{RBL} in next-generation -- $e.g.$ \ac{B5G} and \ac{6G} -- wireless systems applied to \ac{V2X} perception is also discussed, with a focus on its role in standardization efforts and its relevance across automotive and industrial domains.
\end{abstract}

\vspace{-1ex}

\glsresetall
\begin{IEEEkeywords}
\Ac{V2X}, \ac{RBL}, rigid body tracking, \ac{6G}, \ac{ISAC}, wireless systems.
\end{IEEEkeywords}
\glsresetall

\vspace{-2.5ex}
\section{Introduction}
%
%
%
%
%
\vspace{-.5ex}

\Ac{WL} can be regarded as an early example of \ac{ISAC} that demonstrates how communication signals can also be used to extract environmental information ($i.e.$, the location of users) using signals originally intended for communications \cite{LimaAccess2021}.
These capabilities become particularly relevant in certain industry verticals, such as \ac{ITS} and automotive, where there is an increasing need for \ac{V2X} perception and convergence of previously independent functions, such as sensing, communication, and computing.
Indeed, \ac{WL} has been successfully used in a wide range of applications including security, smart homes, industrial automation, robotics, vehicular networks, localization of \acp{UAV}, and more.

In terms of the fundamental concept or metrics for \ac{WL}, $i.e.$, the type of information extracted from wireless signals for the purpose of localization, many \ac{SotA} techniques have been proposed, including, methods based on finger-prints, \ac{RSSI}, \ac{AoA}, radio range, connectivity, hop-counts, etc.

The \ac{WL} literature also exhibits vast diversity in terms of the underlying mathematical approaches used in the design of positioning algorithms, with examples ranging from purely algebraic methods exploiting \ac{MDS}, and its further developed \ac{SMDS} variant, methods based on optimization-theoretical techniques, such as \ac{SDP}, \ac{PoCs}, \ac{FP} and \ac{MC}, to schemes based on \ac{ML}, which gain more popularity in recent years.

Finally, in addition to the aforementioned methods, abundance also exists in terms of offered features, such as robustness against noise, bias, \ac{LOS} and \ac{NLOS} conditions, and mitigation of effects, including not only information scarcity, uncertainty of anchor points, co-existence of near- and far-field waves, but also physical problems, such as missing links for measurements that are needed for the positioning algorithms.
These challenges will be further discussed in Section \ref{sec:chal}.

Despite the healthy breadth of topics covered by the literature, one important {practical aspect} that has not been fully addressed by most of the \ac{SotA} \ac{WL} methods (nor commercial wireless systems, $e.g.$, \ac{5G}) is the fact that, in many use cases, targets could be better represented as \ac{3D} objects.
Aiming to address this matter, a growing literature is emerging, in which each target is modeled not as a single point, but as a group of inter-connected points with a fixed and known arrangement, $i.e$, a rigid body \cite{ChepuriTSP2014, DongTWC2023, Chen_2015, fuehrling2025robustegoisticrigidbody}.
In our view, a rigid body representation of objects would enhance most of the applications mentioned before, such as control, mobility, and safety for autonomous robots, vehicles, and \acp{UAV}, and networks consisting of these wireless devices.

Thanks to the inherent inclusion of target shape and orientation models, which can be either used as prior information or jointly estimated within the solution of the localization problem, these \ac{RBL} techniques, have been shown to provide higher accuracy than \textit{conventional} (point-based) \ac{WL} methods.
Thus, \ac{RBL} expand the notion of a single \textit{reference} position to position, translation, and rotation, which are very valuable information in many realistic scenarios.
In view of the latter, a significant amount of work and schemes for \ac{RBL} have appeared.
%
%
Remarkably, anchorless localization methods will become more relevant in the future, taking into account the fact that sidelink-based positioning has recently been added to the \ac{5G} specifications in 3GPP Release 18, which is a positioning technique purely operating between \acp{UE}.

With the recent popularity of autonomous driving, extending the idea of rigid body localization even further leads to rigid body tracking approaches that can be used for collision detection.
%
%
%
In that sense, in \cite{Chen_2015}, a two-stage approach was used to estimate rotation, translation, angular velocity and translational velocity by range and Doppler measurements, making use of various \ac{WLS} minimizations.

Emphasizing the distinction between \ac{WL}/\acf{ISAC} \cite{LimaAccess2021}, where wireless connectivity is a fundamental component of the technology, and other approaches based on, $e.g.$, inertial sensing 
or machine vision, which are outside of the scope of this article,
it can be said based on the above that with respect to \ac{RBL}, the current focus of the \ac{WL}/\ac{ISAC} research community is essentially extending/applying the multitude of techniques found in the vast \ac{ML} literature to the \ac{RBL} problem,  which could be jointly referred to as \ac{W-RBL}.

The rest of the article is organized as follows:
Section \ref{sec:stand}, provides industry and standardization perspectives on the need for adopting \ac{RBL} features in \ac{B5G} systems, most imminently \ac{6G}, as well as potential use cases and applications.
Section \ref{sec:sys} offers a brief introduction of \acf{RBL}, defines the most relevant mathematical relationships, and discusses the transition from point-based localization to rigid body localization.
Next, in Section \ref{sec:chal}, main applications, challenges and opportunities that come along with the usage of rigid bodies are discussed in detail.
Finally, Section \ref{sec:conc} closes the article with final remarks and future research directions.

\vspace{-2ex}
\section{Rigid Body Localization in beyond-5G and 6G}
\label{sec:stand}

\begin{figure*}[b!]
\centering
\vspace{-2ex}
\includegraphics[width=\textwidth]{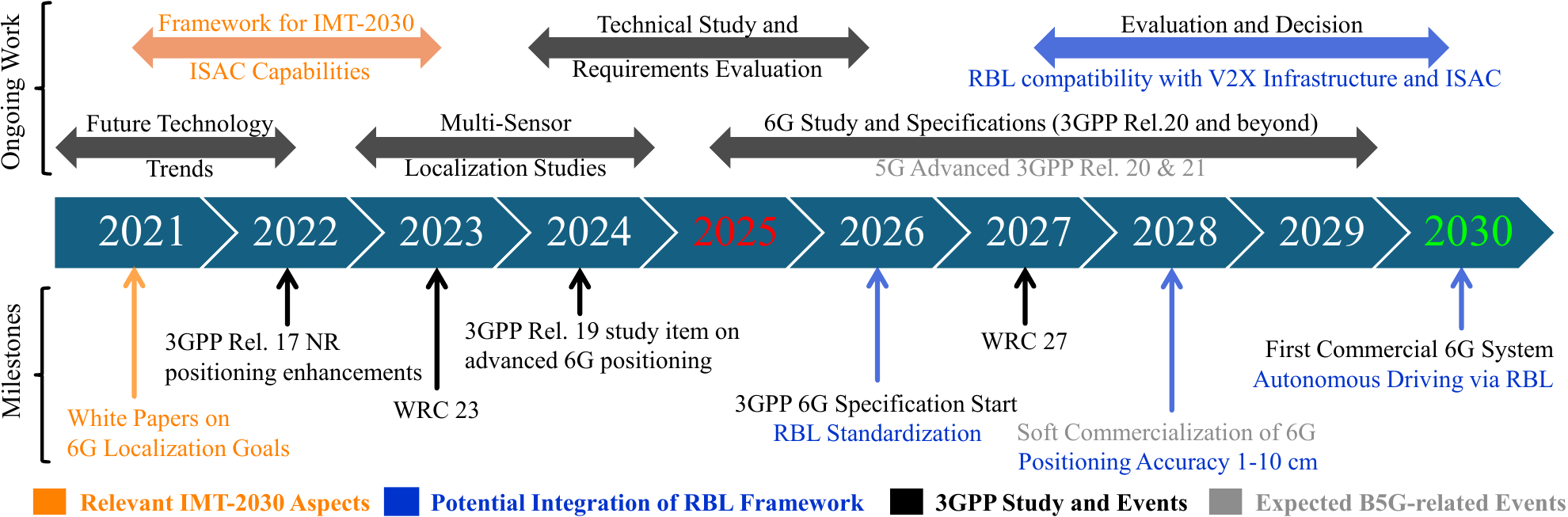}
\caption{Suggested integration of RBL framework within the (pre)standardization of 6G \textit{in-line} with the IMT-2030 use cases and capabilities.}
\label{fig:6G_timeline}
\end{figure*}

Positioning services have been standardized in existing 3GPP commercial networks, such as Long Term Evolution (LTE, 4G) and New Radio (NR, 5G), mainly as a \textit{necessary complement} to conventional Global Navigation Satellite Systems (GNSS), for certain mission critical applications with stringent requirements in terms of positioning accuracy, latency, availability, and/or reliability. In general terms, 5G positioning features include network-based positioning, sidelink-based positioning (without direct use of network signals), and hybrid-enhancements which leverage positioning capabilities in other available radio access technologies (RATs), such as ultra-wide band (UWB), GNSS, Bluetooth, WiFi, etc. A variety of methodologies have been deployed including time difference, round trip time (RTT), carrier phase measurements, angle of arrival/departure (AoA, AoD), etc., and improvements have been made in terms of signaling overhead, coverage scenarios, integrity, accuracy, etc. Since the beginning of 5G-Advanced (5G-A, release 18 and beyond) positioning enhancements based on Artificial Intelligence and Machine Learning (AI/ML) have also been studied and specified. However, a more general framework for RBL has not been considered yet in 5G systems, and hence, it is an interesting (and potentially needed) {positioning expansion/enhancement} to be developed for future commercial systems, such as 6G. 

\vspace{-2ex}
\subsection{{Standardization Perspectives and Applications}}

At the time of writing, the standardization of 6G continues with the recent approval of the new Study Item for 6G Radio~\cite{RP-251881}. This is an important step, in particular towards the definition of the physical layer (e.g., signals, channels) and related procedures and functionalities to support the previously identified use cases and requirements. Indeed, some use cased described in~\cite{22.870} (e.g., environment object reconstruction, network assisted smart transportation, cooperating mobile robots, and real time digital twins) require advanced localization capabilities and can significantly be enhanced by RBL. Thus, it is clear that sensing and advanced positioning/localization are key to support use cases in several verticals, such as ITS, automotive, and automated/smart factories. In general, complex applications requiring localization and tracking, object reconstruction, and environment perception can be benefited from RBL capabilities. In particular, for the automotive sector, RBL is also very \textit{compatible} with the use of (next-generation) \mbox{\ac{V2X} infrastructure \textit{as a sensor}} in connected/digital roads~\cite{15:00020}.
Applications include use cases~\cite{09:00055} requiring real time environmental modeling and/or intelligent automated driving systems, such as automatic valet parking, lane merging/maneuvering, smart intersections, protection of vulnerable road users, and automatic charging/fueling of (electric) vehicles.

\begin{table*}[t]
\centering
\caption{Mapping of RBL properties and features (Sections IV) to relevant 6G standardization considerations (Section II).}
\vspace{0.5ex}
\renewcommand{\arraystretch}{1.3}
\setlength{\tabcolsep}{4pt}
\begin{tabular}{|c|p{4cm}|*{9}{c|}}  
\cline{3-11}
\multicolumn{2}{c|}{} & \multicolumn{9}{c|}{{Fundamental Properties and Features of RBL (Sections IV)}} \\
\cline{3-11}
\multicolumn{2}{c|}{} &
\rotatebox{90}{\parbox{3.4cm}{\centering \bf Algebraic Approaches \\ (IV-A-1)}} &
\rotatebox{90}{\parbox{3.4cm}{\centering \bf Optimization Approaches \\ (IV-A-2)}} &
\rotatebox{90}{\parbox{3.4cm}{\centering \bf Message Passing \\ (IV-A-3)}} &
\rotatebox{90}{\parbox{3.4cm}{\centering \bf Matrix Completion \\ (IV-B)}} &
\rotatebox{90}{\parbox{3.4cm}{\centering \bf Fundamental Limits \\ (IV-C)}} &
\rotatebox{90}{\parbox{3.4cm}{\centering \bf Node Placement \\ (IV-C)}} &
\rotatebox{90}{\parbox{3.4cm}{\centering \bf Egoistic RBL \\ (IV-D-1)}} &
\rotatebox{90}{\parbox{3.4cm}{\centering \bf RB Tracking \\ (IV-D-2)}} &
\rotatebox{90}{\parbox{3.4cm}{\centering \bf Semantic Localization \\ (IV-D-3)}} \\
\hline
\multirow{8}{*}{\rotatebox{90}{\parbox{5.7cm}{\centering {Standardization Considerations for 6G \\ (Section II)}}}} 
& {A. Backward Compatibility with current Standards} &\tablecheck &\tablecheck &\tablecheck &\tablecheck &\tablecheck & &\tablecheck &\tablecheck  &  \\
\cline{2-11}
& {B. Accuracy and High Precision Localization} &\tablecheck &\tablecheck & \tablecheck& & &\tablecheck & &\tablecheck   &\tablecheck    \\
\cline{2-11}
& {C. Energy Efficiency and Hardware Constraints} &\tablecheck   &  &\tablecheck &  &\tablecheck &\tablecheck  &\tablecheck   &  &\tablecheck \\
\cline{2-11}
& {D. Ultra-Low Latency: Implementation Complexity} &\tablecheck &  &\tablecheck &  &\tablecheck  &  &   & &\tablecheck \\
\cline{2-11}
& {E. Standardized Parameter Configuration} &   & &   &   &\tablecheck &\tablecheck &    &   & \tablecheck  \\
\cline{2-11}
& {F. Integration with Emerging Use Cases} &\tablecheck   &\tablecheck &\tablecheck   & \tablecheck   &   &   & \tablecheck& \tablecheck  &   \\
\cline{2-11}
& {G. Robustness to Practical Impairments} &   & &   &\tablecheck   &\tablecheck   &\tablecheck    &  & &   \\
\cline{2-11}
& {H. Testbeds, Trials, and Reference Designs} &     &   &   &   &   & \tablecheck  &  \tablecheck &  \tablecheck & \tablecheck\\
\hline
\end{tabular}
\label{tab:feature_mapping}
\vspace{-2ex}
\end{table*}

\newpage
All in all, current developments indicate that \ac{RBL} could  be one of the most interesting novelties in 6G. 
To emphasize the potential role of \ac{RBL} in future 6G systems, Figure \ref{fig:6G_timeline} provides an overview of past and future milestones, highlighting the capabilities and use cases specified in \mbox{IMT-2030}, also highlighted later on in Table \ref{tab:feature_mapping}.
%


\section{Rigid Body Localization System Model}
\label{sec:sys}

\subsection{{From Wireless Point-based Localization to Wireless Rigid Body Localization (W-RBL)}}

As described before, in the past, various approaches have been proposed for wireless localization that exploit many different types of information, as well as mathematical models to solve the problem.
Whereas single point localization tend to use a single measurement to a set of anchors, the same can be applied to rigid bodies, where each sensor in the rigid body performs a measurement to the set of sensors, or vice versa.
The main difference lies in the fixed conformation of the rigid body that need to be incorporated in the corresponding methods, having obtained information through measurements of all the individual sensors.
Thus, any localization method can effectively be reformulated to incorporate the \ac{RBL} constraints by an appropriate insertion of the \ac{RBL} framework given by equation \eqref{eq:sys_mod}, opening significant directions of improving the existing \ac{SotA} methods.

\vspace{-2ex}
\subsection{{Background of Rigid Body Localization}}
\label{sec:rb}

%
Given a rigid body, the position of the $k$-th node within a rigid body can be described by a \ac{3D} coordinate vector $\mathbf{c}_{k} \in \mathbb{R}^{3 \times 1}$,
%
%
which captures the $x$- and $y$-coordinates of the $k$-th node of the rigid body respectively, for all $K$ points defining the rigid body.

\begin{figure*}
\centering
\begin{subfigure}{\textwidth}
\centering
\vspace{-5ex}
\includegraphics[width=\textwidth]{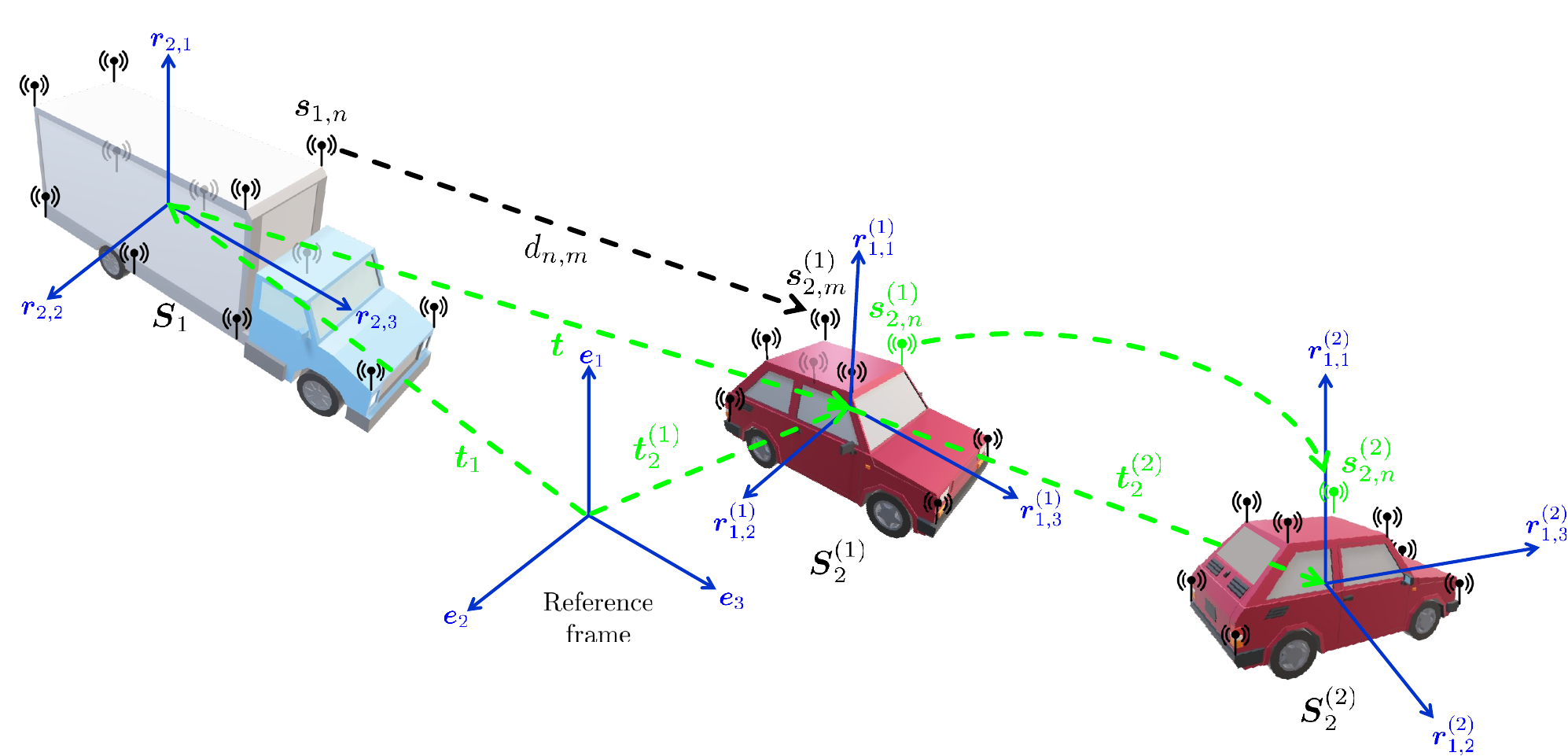}
\caption{\Ac{V2V} Rigid body scenario and rigid body transform for the next timestep.}
\label{fig:subfig1}
\vspace{4ex}
\end{subfigure}
\begin{subfigure}{\textwidth}
\centering
\includegraphics[width=\textwidth]{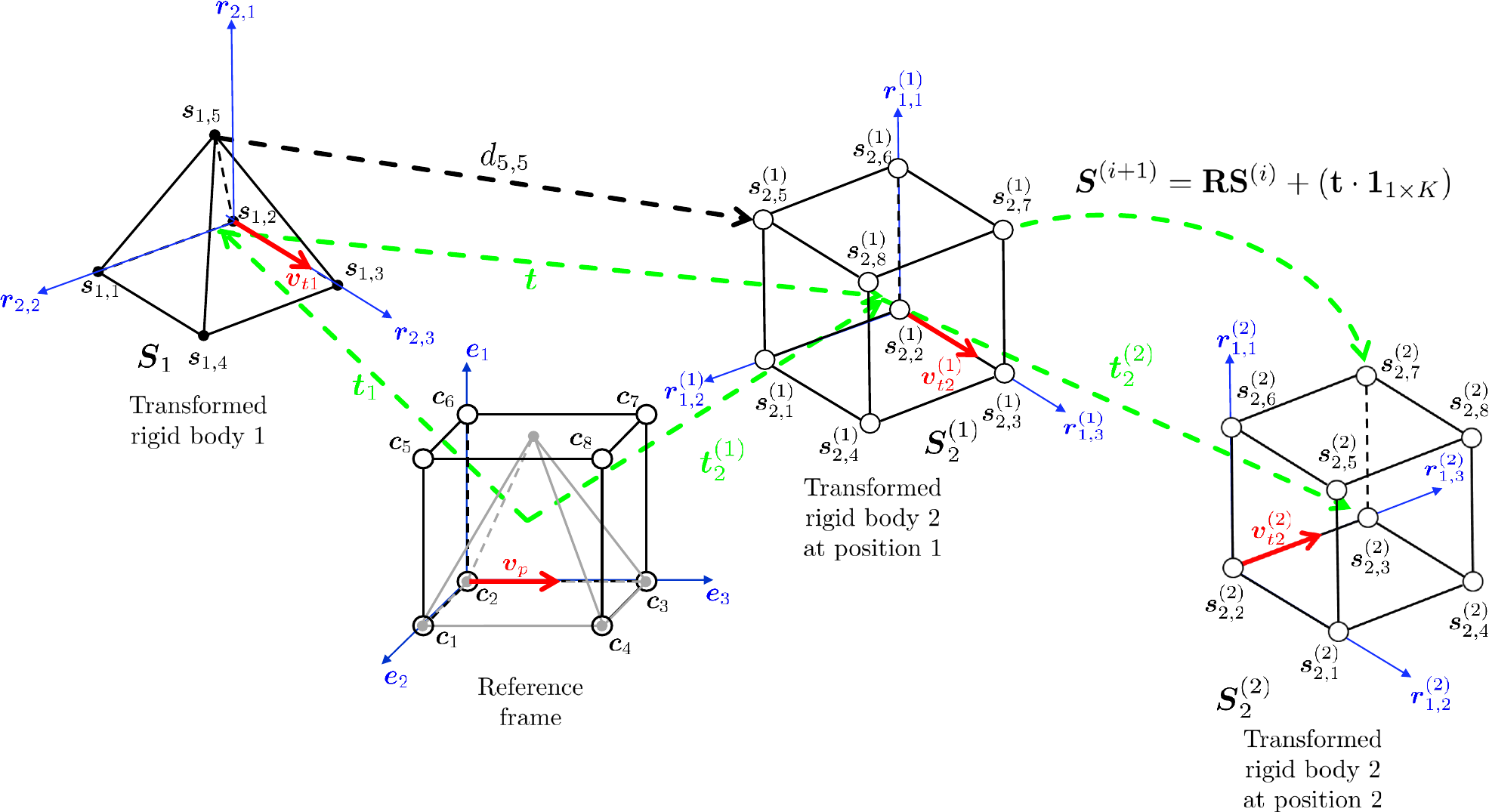}
\caption{Abstract representation of a \ac{RBL} scenario with semantic features.}
\label{fig:subfig2}
\vspace{2ex}
\end{subfigure}
\caption{Illustrations of multiple rigid body scenarios. 
(a) A \ac{V2V} rigid body scenario in which a truck $\boldsymbol{S}_1$ localizes a target rigid body car $\boldsymbol{S}_2$. 
The rigid body transformation is shown by the car at two distinct positions, $\boldsymbol{S}_2^{(1)}$ and $\boldsymbol{S}_2^{(2)}$. 
Without loss of generality, the initial position is represented by the conformation matrix $\boldsymbol{C}$ in the reference frame, defining the rigid body’s shape and orientation. 
The subsequent position $\boldsymbol{S}_2^{(2)}$ is determined according to equation \eqref{eq:sys_mod}, obtained by applying a rotation matrix $\boldsymbol{R}$ and a translation vector $\boldsymbol{t}$ to $\boldsymbol{S}_1^{(1)}$;
%
(b) An abstract representation of the scenario in subfigure (a), where each rigid body has a distinct shape described by its conformation matrix, $\boldsymbol{C}_1$ or $\boldsymbol{C}_2$. 
Also shown is a semantic rigid body model, in which a unit-norm vector $\boldsymbol{v}_P$ points in a predefined direction and is transformed into the target vector $\boldsymbol{v}_T$ via the rigid body transformation, illustrating multiple independent rigid bodies.}
\label{fig:RB_gen}
\end{figure*}

As illustrated in Figure \ref{fig:RB_gen}, following the \ac{SotA} system model of the standard \ac{RBL} framework \cite{ChepuriTSP2014,DongTWC2023,Chen_2015,Nic_RBL,fuehrling20246drigidbodylocalization,fuehrling2025robustegoisticrigidbody,fuehrling2025fundamentallimits,fuehrling2025SMDS,Jiang2019}, a change in position and orientation of a \ac{RBL} can be described by the joint affine transform of the $K$ coordinates of a rigid body, collected in the conformation matrix $\mathbf{C}=[\mathbf{c}_1, \cdots\!, \mathbf{c}_k, \cdots\!, \mathbf{c}_K]\in \mathbb{R}^{3\times K}$, given by \vspace{-1ex}
\begin{equation}
\begin{split}
\mathbf{S} &\triangleq [\mathbf{s}_1, \cdots\!, \mathbf{s}_k, \cdots\!, \mathbf{s}_K] \\&= \mathbf{R} \!\cdot\! [\mathbf{c}_1, \cdots\!, \mathbf{c}_k, \cdots\!, \mathbf{c}_K] +\;\! [\mathbf{t}, \cdots\!, \mathbf{t}, \cdots\!, \mathbf{t}] \\&=\mathbf{R}\mathbf{C}+ \mathbf{t}\cdot\mathbf{1}_{1\times K} , 
\end{split}
\label{eq:sys_mod}
\end{equation}
where $\mathbf{S} \in \mathbb{R}^{3 \times K}$ consists of the new coordinates of the rigid body nodes after the transform, $\mathbf{R}\in \mathbb{R}^{3 \times 3}$ and $\mathbf{t} \in \mathbb{R}^{3 \times 1}$ respectively describe the rotation and translation equally to the $K$ original coordinates.

In addition, another important property of rigid bodies is the fixed relative Euclidean distances between the $K$ nodes in $\mathbf{C}$, which must also be satisfied in $\mathbf{S}$.
Therefore, such information is incorporated in \ac{RBL} methods as a necessary constraint.

The implication for \ac{RBL} is that the known structure of the rigid body in $\mathbf{C}$ will be exploited to aid the estimation of the localized positions in $\mathbf{S}$, where the affine transform $\mathbf{R}$ and $\mathbf{t}$ is unknown, and only some noisy observation of $\mathbf{S}$, or other kinds of measurements related to $\mathbf{S}$ are available.


\section{Algorithms, Challenges and Applications}
\label{sec:chal}

With the main idea of rigid bodies and a clear mathematical description defined, this chapter aims to discuss various approaches, problems and solutions to \ac{RBL}, drawing an analogy to single-point localization schemes and presenting which problems have been solved for those schemes, but are open to investigate for the special case of rigid body targets.

This includes the discussion for multiple different localization approaches, certain challenges that can occur in \ac{RBL}, but also conventional wireless localization problems, as well as some insights into next-generation extensions of \ac{RBL}.

\subsection{{Localization Approaches}}

\subsubsection{{Algebraic Approaches}}

\begin{figure*}[b]
\centering
\includegraphics[width=\textwidth]{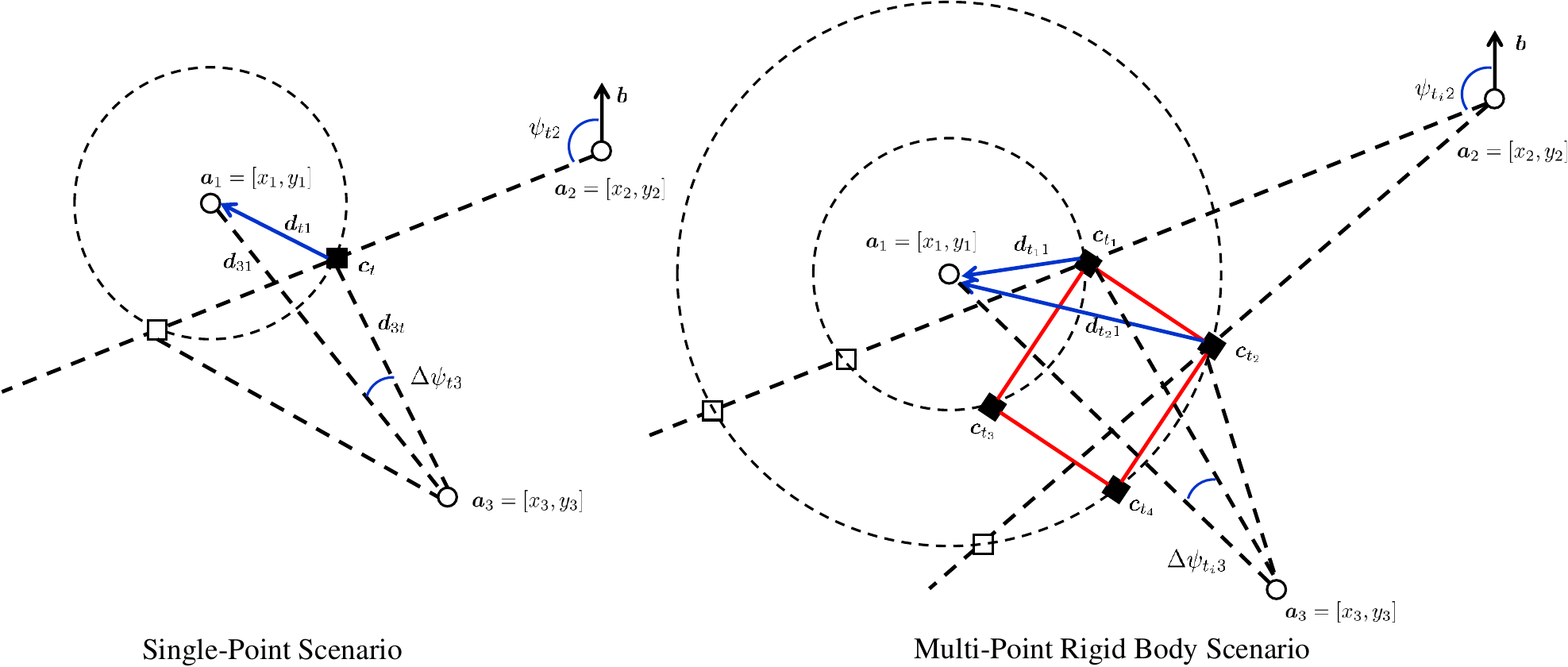}
\caption{Combination of different measurement types for single-point localization and multi-point rigid body (red lines) localization, from anchors $\boldsymbol{a}_i$ (white balls) to sensor $\boldsymbol{c}_{t_i}$ (black square), for algebraic localization, exploitable for all approaches.}
\label{fig:MeasComb}
\vspace{2ex}
\end{figure*}

Inspired by the aforementioned discussion, it can be considered what type of measurements can be used and combined for localization, which measurements are the most useful, and how one can benefit from the known shape of the rigid body.
Combining all these measurements leads to different algebraic and geometric approaches resulting in diverse estimation methods, with the most well known examples being triangulation, \ac{MDS} \cite{fuehrling2025robustegoisticrigidbody,Nic_RBL}, and \ac{SMDS} approaches \cite{fuehrling2025SMDS}.
Therefore, to give an insight on how these methods can be adapted to rigid bodies, Fig. \ref{fig:MeasComb} illustrates how different type of measurements can be combined for single-point localization, as well as \ac{RBL}.
On the left hand side, a single target is localized via range, \ac{AoA} and \ac{ADoA} measurements, whereby combination of the later the position can be exactly defined.
On the right hand side, a multi-point rigid body is localized by the combination of the same measurements.
What can be observed is that due to the known shape of the rigid body one might be able to save certain measurements, since the conformation matrix adds a constraint that helps in the localization process.
Additionally, it can be seen that for a rigid body, angle measurements are much more relevant than range measurements, since two points can have the same distance to an anchor, as illustrated, while the angles change, which was investigated in \cite{fuehrling2025SMDS}.
However, which kind of measurements are the most effective and how to process them is still open to investigate and only touched slightly in \cite{fuehrling2025fundamentallimits}, which investigates the fundamental limits of \ac{RBL}, as discussed in Section \ref{sec:fund}.

\vspace{1ex}
\subsubsection{{Optimization Approaches}}
In terms of optimization, many well-known \ac{SotA} methods have been applied to the \ac{RBL} problem, including methods such as least squares, \ac{SDP}, \ac{PoCs}, \ac{FP} and \ac{MC} schemes have been deployed \cite{ChepuriTSP2014,DongTWC2023,Chen_2015,fuehrling2025robustegoisticrigidbody,Jiang2019}.
In contrast to the pure algebraic approaches, optimization based methods tend to describe the desired problem mathematically in a way such that through common techniques, such as gradient descent methods, an optimization problem dealing with the main objective, as well as, for example, noise constraints can be solved.
Depending on the problem statement itself, these approaches can lead to efficient estimations, possibly leading to closed-form solutions, which can also be adapted to rigid body frameworks.

\vspace{1ex}
\subsubsection{{Message Passing}}

The algebraic and classical optimization methods described in the previous sections, while providing optimal solutions with possibly even closed-form expressions, typically require high complexity and may present numerical instability due to the inherent matrix inversions, and also exhibit low robustness to additional/incomplete information in the system.
As an alternative to the conventional matrix inversion-based approaches, \ac{MP} algorithms can offer considerable robustness in both formulation and implementation, hence unsurprisingly, have been widely adopted across diverse domains of signal processing and wireless system design, including channel coding, distributed estimation, and \ac{ML}. 

\Acl{MP} methods operate on a factor graph representation, where each node corresponds to either an observed quantity or a latent variable, and edges indicate statistical dependencies, and under suitable conditions, are capable of achieving Bayes-optimal solutions. 
The key idea lies in computing and exchanging locally derived messages, typically sufficient statistics or parametric forms of probability distributions, between nodes.

\begin{figure}[H]
\centering
\includegraphics[width=\columnwidth]{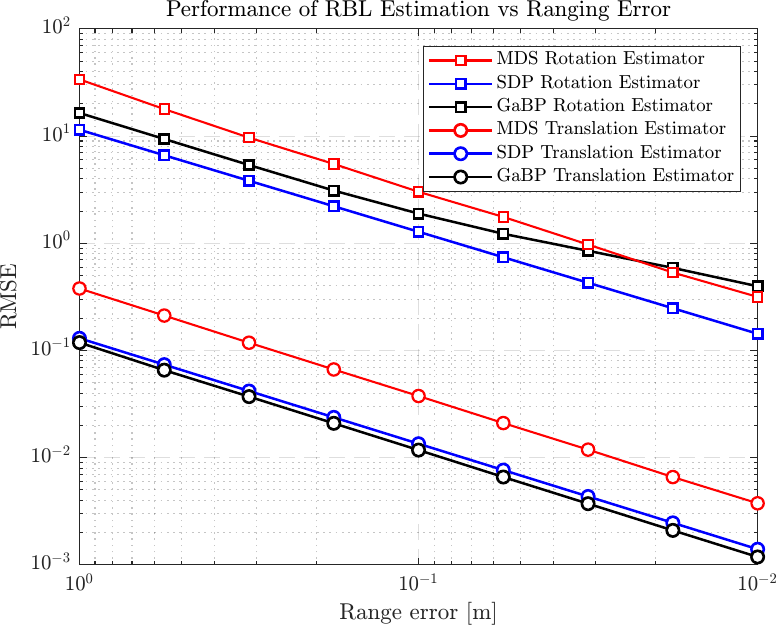}
\vspace{-3ex}
\caption{\Ac{RMSE} (in meter for translation and in degrees for rotation) of distance measurement-based rigid body parameter estimation via \ac{MDS}, \ac{SDP} and \ac{GaBP}. Simulation setup follows \cite{fuehrling20246drigidbodylocalization}, where the target rigid body is modeled as a cube with 8 nodes placed at the corners, and the anchors are placed at the corners of another larger cube.}
\label{fig:appraoaches}
\end{figure}

Through iterative updates, the algorithm converges to consistent marginal estimates of the unknown variables.
Given the above, building upon the system model in equation \eqref{eq:sys_mod}, tailored \ac{MP} algorithms can be developed to solve the \ac{RBL} problem of jointly estimating the \ac{3D} rotation matrix and translation vector \cite{fuehrling20246drigidbodylocalization}.
Specifically, the distributed nature makes \ac{MP} especially well suited for \ac{RBL} scenarios involving incomplete or noisy measurements, with its inherent adaptability allowing for robust estimation in systems with structural irregularities or missing data.
Furthermore, the probabilistic foundation of \ac{MP} aligns well with soft-decision localization, enabling the characterization of estimation uncertainty via distributions or confidence ellipses. 
These properties, along with robustness to missing observations and given proper algorithm design to prevent convergence issues, such as self-interference or divergence due to loops or misconfigured messages, position \ac{MP}-based approaches as a promising solution for future wireless \ac{RBL} systems as also elaborated in the following section.
To conclude, it has been illustrated in Figure \ref{fig:appraoaches} how the different approaches can perform under varying range error for the translation and rotation estimation.

\vspace{-2ex}

\subsection{{Incomplete Observations}}
\label{sec:MC}

\begin{figure}[t]
\centering
\includegraphics[width=\columnwidth]{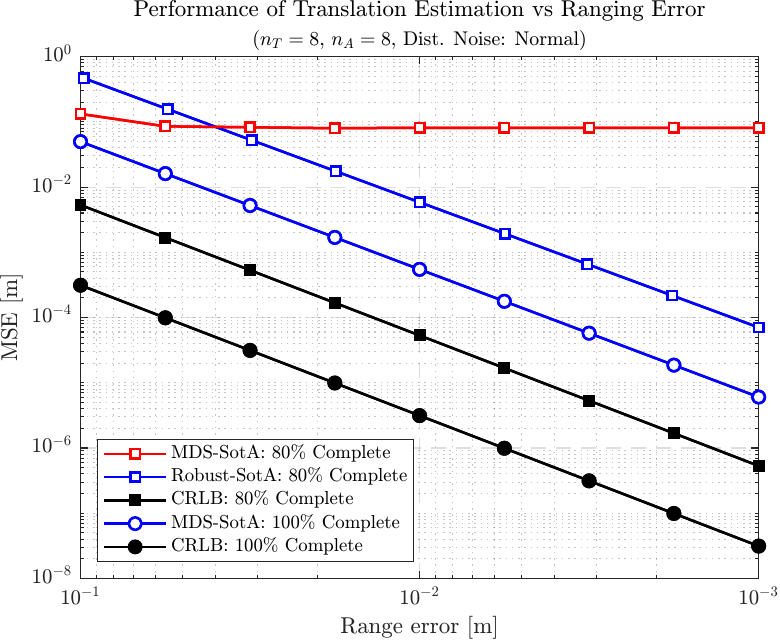}
\vspace{-2ex}
\caption{\Ac{MSE} translation estimation performance for systems with complete and incomplete information compared to the \ac{CRLB}. The simulation setup follows \cite{fuehrling2025robustegoisticrigidbody}, where the target rigid body is modeled as a car, while the anchors are modeled as a second rigid body, depicted as a truck, as shown in Figure \ref{fig:subfig1}.}
\label{fig:MC}
\vspace{-3ex}
\end{figure}

As hinted by the prior sections, independent of the metric used for localization, a common issue regularly occurring is the observation of incomplete measurements, for instance due to blocked paths.
Just to name one example, when considering range measurements from anchors to sensors, a \ac{LOS} path might be blocked due to the structure of the rigid body, which leads to zeros occurring in the observed distance matrix. 
In the case of range measurements these zeros are especially harming, since putting a zero is equivalent to stating that the measured distance is exactly zero, which poisons the algorithms.
Therefore, these zeros have to be handled with care, since either the algorithm has to be adjusted to be aware of the missing information, or something has to be done to reconstruct the lost data.

Since this is a common problem, there is a vast literature that designs robust frameworks by leveraging methods, such as \ac{ML}, data fusion, graph-based methods, or matrix completion.
One possible solution to this problem, which highly benefits from the structure of the rigid body, captured in the conformation matrix $\bm{C}$ are \acf{MC} solutions, which has been proposed in \cite{fuehrling2025robustegoisticrigidbody}.
\ac{MC} literature offers a wide range of different methods that can be used for completion, such as Nystr\"om approximation, \ac{SDP} solutions, proximal gradient methods, and even discrete aware algorithms, often used in scenarios such as recommender systems in computer science, wave channel estimation in wireless communication, but also localization algorithms in signal processing.
By knowing the conformation structure of the rigid body itself, as well as the positions of the anchors, it is possible to construct a hollow squared \acf{EDM}, containing the intra-distances of the anchors to one another, the intra-distances of the points in the rigid body itself, as well as the incomplete measured cross-distances.
This information alone is sufficient to complete the matrix with high accuracy to improve the performance significantly.

Recent contributions have used such matrix completion methods, such as the one in \cite{fuehrling2025robustegoisticrigidbody}, as illustrated in  Figure \ref{fig:MC}, which shows that the use of such completion techniques can significantly improve the performance compared to other conventional method, in case that the observations are incomplete.
Nevertheless, it is also shown that the estimate is not very close to the \ac{CRLB}, which indicates that there is still room for improvement.

\subsection{{Fundamental Limits}}
\label{sec:fund}

Another important aspect in terms of real life scenarios and industrial standards are the fundamental limits of estimators, as indicated in the previous sections, which partially boils down to the sensor deployment on the rigid body itself, $i.e.$, how many sensors to use and where to put them on a specific vehicle?
Recently, \cite{fuehrling2025fundamentallimits} has investigated the fundamental limits of \ac{RBL} in terms of the \ac{CRLB}, which is a lower bound on the variance of any unbiased estimator, and can be used to evaluate the performance of an estimator, based on the type of measurement and error distribution, as shown already in Figure \ref{fig:MC}.

While \cite{fuehrling2025fundamentallimits} offers a generic framework to construct the \ac{CRLB} for various \ac{RBL} scenarios, it is left to investigate how the \ac{CRLB} can be used to optimize the sensor deployment on the rigid body itself.
As an example, take the scenario presented in Section \ref{sec:MC}, where some measurements might be missing due to \ac{NLOS} paths, where the sensors should be positioned, such that the amount of possible \ac{NLOS} paths is minimized.
To that extend, if the sensors will connect to anchors positioned high up in road side units it might be useful to place the sensors on top of the rigid body, while, in \ac{V2V} scenarios, it is advisable to place them in the front and the back of the vehicles.

For conventional single-point localization scenarios, multiple \ac{SotA} methods have investigated the problem of optimal anchor placement, as shown in \cite{zhao2013optimal} using frame theory, however, for rigid bodies, the problem is not as straightforward.
To summarize, it was found that for single targets the anchors can be placed to minimize the frame potential, $i.e.$, how closely a set of anchors in a Hilbert space approximates an orthonormal basis.
For multiple targets, the anchors can be placed sequentially optimizing the position of the anchor with respect to each of the targets position and the corresponding \ac{CRLB}, which is build on the problem of single-target localization using \ac{AoA} measurements.
As a result, \cite{zhao2013optimal} proposed the optimal anchor placements in 2D and 3D for different amount of anchors, which can be translated to rigid bodies, illustrating the difficulty of placing two sensors only for rigid bodies of higher dimensions, where for a 2D rigid body the anchors need to be placed in 3D space to be optimal, however, for a 3D object this is not possible anymore.

With a framework for fundamental limits in hand, and the work that has been done, it is left to do further research and investigations on how the presented concepts can be applied to \ac{RBL}, in order to define the best possible anchor and sensor placement, as well as the best type of information to measure, similar to the multi target scenario.
Similarly, the effect of sensor uncertainty of the anchors, as well as the sensors on the rigid body itself need to be studied in more depth.


\vspace{-2ex}
\subsection{{Towards Next-Gen Rigid Body Localization and Tracking}}

\subsubsection{{Anchorless Egoistic Rigid Body Localization}}

Another more recent relevant approach in terms of autonomous vehicles proposes a relative egoistic rigid body localization method in an anchorless scenario \cite{fuehrling2025robustegoisticrigidbody,Nic_RBL}, estimating the relative translation and rotation between two rigid bodies by measuring the cross-body \ac{LOS} distances between the bodies, which is visualized in Fig. \ref{fig:subfig1}.
This relative rigid body localization is especially useful in scenarios where there are no, or only a few anchors deployed, such that the bodies can find the relative translation and rotation even without any external sensors, which can be useful for collision detection for vehicles or \acp{UAV} or for other autonomous driving scenarios, only to name a few possible use-cases.

However, there exists one {major flaw} related to a real life implementation of such a system, which is the following.
As explained in equation \eqref{eq:sys_mod} and Section \ref{sec:sys}, each rigid body has a known conformation matrix $\boldsymbol{C}$.
The problem occurs when localization methods need the conformation matrices of all vehicles in order to work, which is not the case in real life, since a body can indeed measure the distances to other sensors, but does not necessarily know the other bodies structure.
To solve such issues for real life applications regulations on how knowledge on $\boldsymbol{C}_i$ should be acquired and distributed would be required.
Otherwise, egoistic methods need to and are currently be developed, which are able to perform localization purely by measurements, without any prior knowledge about the target rigid body itself, such as the ones proposed in \cite{fuehrling2025robustegoisticrigidbody,Nic_RBL}.


\vspace{2ex}
\subsubsection{{From Rigid Body Localization to Rigid Body Tracking}}

Inspired by the aforementioned work and as a logical next step towards autonomous driving, one has to extend the concept from rigid body localization to rigid body tracking.
While for single-point based localization methods it is more intuitive to connect the model to time, adding a time constant to the rigid body system model can become a more complicated task.
Single-point based systems can easily incorporate time by applying a velocity to their point, which is not directly possible  for rigid bodies.
In terms of rigid body tracking, substantial research has been done in the field of visual tracking, for example tracking robots, faces or general gestures by cameras.
In that sense many approaches, such as feature based tracking, edge based tracking, or sensor fusion has been proposed, which performs very well for its applications.
Nevertheless, as described before, we are focusing on \acf{W-RBL} and therefore have to search for a different approach.
What can be done to model the motion of the entire body instead of visually capturing its motion is to add an angular velocity and a translational velocity \cite{Chen_2015,fuehrling20246drigidbodylocalization}.
Mathematically this can be described in simple form per sensor as

\begin{equation}
\dot{\boldsymbol{s}}_i=[\boldsymbol{\omega}]^\times\boldsymbol{R}\boldsymbol{c}_i+\dot{\boldsymbol{t}}_i, 
\label{eq:track}
\end{equation}
where $\dot{\boldsymbol{s}}_i$ denotes the velocity of the individual sensor in the inertial frame, $[\boldsymbol{\omega}]^\times$ being the cross product operator matrix applied to the angular velocity and $\dot{\boldsymbol{t}}_i$ being the translational velocity.

Another important aspect to keep in mind is what type of tracking to pursuit, since multiple approaches are known and can be called tracking.
The most common tracking approach is to a notion of time in the system model, which coherently influencing the velocity or similar parameters.
These type of tracking applications are commonly solved via \ac{SotA} methods, such as Kalman filters.
Another more relevant type of tracking with respect to rigid body applications, followed in \cite{Chen_2015} and \cite{fuehrling20246drigidbodylocalization}, is to look at moving rigid bodies in time from frame to frame, only working at one time instance at a time, such as shown in equation \eqref{eq:track}.
This type of tracking can be used especially effective in frameworks such as platooning problems.


\subsubsection{{Semantic Localization}}

Finally, as proposed in \cite{fuehrling2025robustegoisticrigidbody} and illustrated in Figure \ref{fig:subfig2} in some applications one cannot distinguish between multiple rigid bodies, which have the same translation and rotation, such that a semantic rigid body model needs to be introduced.
To that extend, as shown in Figure \ref{fig:subfig2}, a rigid body can be modeled by a unit norm vector $v_p$ pointing in a pre-defined direction, which after the rigid body transformation is given by the target vector $\boldsymbol{v}_T=\boldsymbol{Q}\cdot\boldsymbol{v}_P\raisebox{-2pt}{$\big|_{\boldsymbol{t}}$}$, instead of the full conformation matrix $\boldsymbol{C}$ tha tcontains all the information about the rigid body.
This leads to a semantic rigid body localization problem, for which specific semantic algorithms need to be developed for future applications.


\section{Conclusion and Outlook} 
\label{sec:conc}

In this work we have provided a detailed discussion about the usage and benefits of \acf{RBL}, which is a highly promising technique for future \ac{V2X} perception and autonomous driving applications.
In that sense, we have presented the general system model of a rigid body framework, with the transition from a single point localization to \ac{RBL}, enabling the estimation of the orientation of the target instead of only the position.
Furthermore, algorithms, challenges in a vast array of fields and applications have been discussed. 
The discussion about incomplete measurements directly connects to the fundamental limits and sensor deployment, which described how the placement of nodes can influence the performance, leading to conditions where incomplete measurements occur which can be addressed by different methods, such as matrix completion.
Further, it was discussed how modern signal processing methods, including message passing can be redesigned to fit the \ac{RBL} problem, resulting in significant complexity reductions/gains.

Even more relevant for future applications, especially for autonomous driving in rural areas where not many anchors are available, we discussed how multiple rigid bodies can perform localization in anchorless scenarios with respect to each other
and how the whole framework of \ac{RBL} can be extended to rigid body tracking and semantic localization, where the rigid body can be localized only by its translation and rotation without the need of a conformation matrix.

Finally, as discussed in Section \ref{sec:stand} and illustrated in Figure \ref{fig:6G_timeline}, positioning applications have been standardized in all existing 3GPP commercial networks, where advanced positioning has been set as a goal for 6G, which offers great potential for future \ac{RBL} applications. 

\newpage


\bibliographystyle{IEEEtran}

\vspace{-9ex}
\begin{IEEEbiography}{Niclas F\"uhrling}
received the B.Sc. degree in electrical and computer engineering from Jacobs University Bremen, Bremen, Germany in 2022, and the M.Sc. degree in electrical engineering with the University of Bremen in 2024. He is currently pursuing his Ph.D. degree in Electrical Engineering at Constructor University (previously Jacobs University), working on a research project focusing on 6G connectivity. His current research interests are wireless communications, signal processing, and positioning.
\end{IEEEbiography}
\vspace{-9ex}
\begin{IEEEbiography}{Hyeon Seok Rou}
(S'19-M'24) is a postdoctoral research associate and lecturer at Constructor University, Bremen, Germany, where he also received the Ph.D. and B.Sc. degree in Electrical Engineering, respectively in 2024, and 2021. His research interests lie in the fields of next-generation waveform design, including AFDM, integrated sensing and communications (ISAC), Bayesian statistics, multi-dimensional modulation schemes, and quantum-accelerated optimization techniques for wireless system design.
\end{IEEEbiography}
\vspace{-9ex}
\begin{IEEEbiography}{Giuseppe Thadeu Freitas de Abreu}
[SM’09] is a Professor of Electrical Engineering in the School of Computer Science and Engineering at Constructor University Bremen, Germany. His research interests include communications theory, statistical modeling, wireless localization, wireless security, MIMO systems, ultrawideband and millimeter wave communications, full-duplex and cognitive radio, compressive sensing, connected vehicles networks, and many other topics.
He is serving as an editor to the IEEE Signal Processing Letters and to the IEEE Open Journal of the Communications Society.
\end{IEEEbiography}
\vspace{-8ex}
\begin{IEEEbiography}{David González G.}
 PhD. (S’06–M’15–SM’18) holds a Master’s in Mobile Communications and a Ph.D. in Signal Theory and Communications from the Universitat Polit\`ecnica de Catalunya, Spain. He was a postdoctoral fellow at Aalto University, Finland (2014–2017), and a Research Engineer at Panasonic R\&D Center, Germany. Since 2018, he has been with Continental Automotive Technologies, Germany, leading research on V2X, ISAC, and automotive 5G-Advanced/6G, while participating in 3GPP RAN1, 5GAA, and ETSI ISG ISAC.
\end{IEEEbiography}
\vspace{-80ex}
\begin{IEEEbiography}{Gonzalo Seco-Granados}
(Fellow, IEEE) received the Ph.D. in telecommunications engineering from the Univ. Polit\`ecnica de Catalunya, Spain, in 2000, and an M.B.A. from IESE Business School in 2002. He worked at the European Space Agency on the Galileo system and is now Professor at the Univ. Aut\`onoma de Barcelona. His research focuses on GNSS and 5G/6G localization and sensing. Since 2019, he has been President of the Spanish IEEE Aerospace \& Electronic Systems Society Chapter.
\end{IEEEbiography}

\vspace{-80ex}
\begin{IEEEbiography}{Osvaldo Gonsa}
received the Ph.D. degree in electrical and computer engineering from Yokohama National University, Japan, in 1999, and the M.B.A. degree from the Kempten School of Business, Germany, in 2012. He is currently the Head of the Wireless Communications Technologies Group, Continental AG, Frankfurt, Germany. And since 2020 also serves as a member for the GSMA Advisory Board for automotive and the 6GKom Project of the German Federal Ministry of Education and Research.
\end{IEEEbiography}

\end{document}